\documentclass[11pt,twoside,letterpaper]{article} 
\usepackage{times,fancyhdr}
\usepackage[dvips]{graphicx}

\def\figurename{Figure}
\makeatletter
\renewcommand{\fnum@figure}[1]{\figurename~\thefigure.}
\makeatother

\def\tablename{Table}
\makeatletter
\renewcommand{\fnum@table}[1]{\bfseries\tablename~\thetable.}
\makeatother

\begin{document}
\title{
{\begin{flushleft}
\vskip 0.45in
\end{flushleft}
\vskip 0.45in
\bfseries\scshape Scale Invariant Avalanches:\\ A Critical Confusion}}
\author{\bfseries\itshape Osvanny Ramos\thanks{E-mail: osvanny.ramos@univ-lyon1.fr}\\
Laboratoire PMMH, ESPCI, CNRS UMR 7636,\\
10 rue Vauquelin, 75231 Paris Cedex 05, France.\\
Laboratoire PMCN, Universit\'{e} Lyon 1, CNRS UMR 5586,\\
43 Bld. du 11 novembre 1918, 69622 Villeurbanne, France}

\maketitle \thispagestyle{empty} \setcounter{page}{1}


\begin{abstract}

The ``Self-organized criticality" (SOC), introduced in 1987 by Bak, Tang and Wiesenfeld, was an attempt to explain
the $1/f$ noise, but it rapidly evolved towards a more ambitious
scope: explaining scale invariant avalanches. In two decades, phenomena as diverse as earthquakes, granular piles, snow avalanches, solar flares,
superconducting vortices, sub-critical fracture, evolution, and even stock market
crashes have been reported to evolve through scale invariant avalanches. The theory, based
on the key axiom that a critical state is an attractor of the
dynamics, presented an exponent close to $-1$ (in two dimensions)
for the power-law distribution of avalanche sizes. However, the
majority of real phenomena classified as SOC present smaller
exponents, i.e., larger absolute values of negative exponents, a
situation that has provoked a lot of confusion in the field of
scale invariant avalanches. The main goal of this chapter is to shed light on this issue. The essential role of the exponent
value of the power-law distribution of avalanche sizes is discussed. The exponent value controls the ratio of small
and large events, the energy balance --required for stationary
systems-- and the critical properties of the dynamics. A condition
of criticality is introduced. As the exponent value decreases,
there is a decrease of the critical properties, and consequently
the system becomes, in principle, predictable. Prediction of scale
invariant avalanches in both experiments and simulations are presented.
Other sources of confusion as the use of logarithmic scales, and the
avalanche dynamics in well established critical systems, are also
revised; as well as the influence of dissipation and disorder in
the ``self-organization" of scale invariant dynamics.


\vspace{11pt} \noindent \textbf{Keywords:} avalanches, scale invariance, Self-organized Criticality, avalanche prediction.
\end{abstract}

\section{Introduction}

\begin{quote}
\begin{flushright}
{\it ``But he hasn't got anything on," a little child said.}\\
Hans Christian Andersen in {\it The Emperor's New Clothes}
\end{flushright}
\end{quote}

Scale invariance pervades nature, both in space and time. In space,
it is revealed through the ubiquity of fractal structures; and in
time, with the presence of scale invariant avalanches. Avalanches
can be seen as sudden liberations of energy which has been
accumulated very slowly\footnote{A more general definition of
avalanches is introduced in section~\ref{ising model}.}; and
phenomena as diverse as earthquakes \cite{OFC 1992, Bak and Tang
1989, Jagla 2010}, granular piles \cite{Held et al 1990, Frette et
al 1996, Altshuler et al 2001, Aegerter et al 2003, Costello et al 2003, Nerone et al
2003, Aegerter et al 2004, Aegerter et al 2004, Ramos et al 2009},
snow avalanches \cite{Birkeland and Landry 2002}, solar flares
\cite{Hamon et al 2002}, superconducting vortices \cite{Bassler and
Paczuski 1998, Altshuler and Johansen 2004, Altshuler et al 2004,
Altshuler et al 2004b}, sub-critical fracture \cite{Santucci et
al 2004}, evolution \cite{Sneppen et al 1995}, and even stock
market crashes \cite{Lee et al 1998} have been reported to evolve
through scale invariant avalanches. The signature of the scale
invariance corresponds to a power-law in the distribution of
avalanche sizes, however, the exponents of the power-laws present
in general different values. In 1987, Per Bak and co-workers
introduced the ``Self-organized criticality" (SOC) as an
explanation of scale invariance in nature \cite{BTW 1987}. The SOC
proposes a mapping between scale invariant avalanches and critical
phenomena, with the key axiom that the critical state is an attractor
of the dynamics, provoking the self-organization of the system
towards a critical state \cite{Jensen 1998, Bak 1997}.

\pagestyle{fancy}
\fancyhead{}
\fancyhead[EC]{O. Ramos}
\fancyhead[EL,OR]{\thepage}
\fancyhead[OC]{Scale Invariant Avalanches: A Critical Confusion }
\fancyfoot{}
\renewcommand\headrulewidth{0.5pt}

However, the axiomatic manner in which the base of the theory was
introduced, set SOC as a theory to be proved more than as a theory
to develop. Many theoretical studies focused on mapping SOC into
the formalism of critical points \cite{Alastrom 1988, Pietronero
et al 1994, Vespignani and Zapperi 1997}; others, on developing
models displaying SOC behavior \cite{OFC 1992, Bassler and Paczuski
1998, Sneppen et al 1995}, increasing the members of the SOC family.
However, the number of experiments were rather small, and they focused
mainly on validating the theory, where the main goal was to find
power-law distributions of avalanches \cite{Held et al 1990, Frette
et al 1996, Altshuler et al 2001, Aegerter et al 2003}. The original
work presented an exponent close to $-1$ (in two dimensions), while
many of the experimental and numerical results displayed smaller
values, i.e., larger absolute values of negative exponents.
Regardless this difference, they were classified as SOC, having as a
``heritage" all the critical properties of the original model, thus
bringing a lot of confusion to the field of scale invariant
avalanches. The main goal of this chapter is to shed light on
this issue.

The essential role of the exponent of the power-law in the dynamics
is the first subject of discussion. The exponent value controls the
ratio of small and large events, the energy balance --required for
stationary systems-- and the critical properties of the dynamics.
The causes and consequences of a logarithmic scale, which is a source
of confusion affecting the distribution of earthquakes, are also
discussed in this first part of the chapter. The second part
corresponds to the analysis of avalanches in a well established
critical system: the Ising model. In the third part, the study focuses
on the critical properties of scale invariant avalanches, where a
condition of criticality is introduced. In phenomena evolving through
power-law distributed avalanches, a critical behavior leads to the
unpredictability of the dynamics \cite{Bak 1997}. However, as the
exponent of the power-law decreases, there is a decrease of the
critical properties, and consequently the system becomes, in principle,
predictable. Prediction of scale invariant avalanches in both
experiments and simulations are presented in the fourth part of the
chapter. In the last part, the influence of dissipation and disorder
in the ``self-organization" of scale invariant dynamics is also discussed.

\section{Classification of Scale Invariant Avalanches}\label{classification}

\subsection{Fractals and scale invariant avalanches: the role of the exponent value}

\begin{figure}[b!]
\begin{center}
\vspace{-0.25 cm}
\includegraphics[height=2.071in, width=5in]{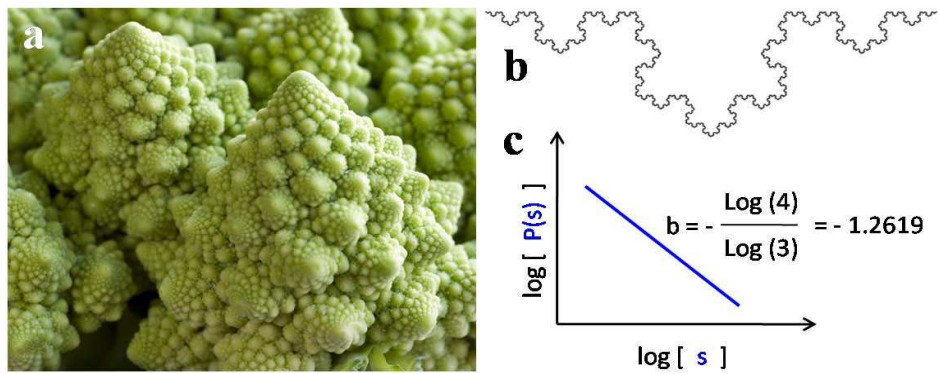}
\end{center}
\vspace{-0.5cm}
\caption{\label{fig:romanesco} (a) Romanesco
broccoli, an example of spatial scale invariance in nature (photo: O. Ramos). (b) The
Koch curve as a representation of an intersection between a plane
and the surface of the broccoli. (c) Scheme of the size distribution
of the different parts in the structure of the Koch curve.}
\end{figure}

The introduction of the fractal dimension by Beno\^{i}t Mandelbrot
in 1975 \cite{Mandelbrot 1975, Mandelbrot 1982} changed the way
nature is perceived; and self-similar branched and rough structures
became more intuitive and natural than the artificially smooth
objects of traditional Euclidean geometry. The self-affine structure
of a Romanesco broccoli (figure \ref{fig:romanesco}a) is an
eloquent example of a natural fractal and if a tiny insect traverses
the vegetable following a straight line, it will surprisingly find a
very long route. The Koch curve \cite{Koch 1904} is a good
representation of this path (figure \ref{fig:romanesco}b), and the
distribution of sizes of its different self-similar parts, resulting
from a triangular ``bending" of the central part of every line, follows a
power-law $P(s)\sim s^b$ with an exponent $b=-1.2619$
\cite{Mandelbrot 1982} (figure \ref{fig:romanesco}c). A similar
analysis over a smooth path donates an exponent equal -1. The
absolute value of the exponent characterizes the trajectory and
provides its dimension, which is fractal in the case of the
broccoli. This fractal dimension is the main concept of the theory
introduced by Mandelbrot; and by using the approach of ``filling the
space", it can be understood rather intuitively: a smooth line
``fits" in one dimension, while the rougher the self-affine curve
(the higher the fractal dimension), the closer it is to fill a
two-dimensional space. For the same reason, self-affine surfaces, as
the one of the broccoli, present fractal dimensions between 2 and 3.

Through the fractal dimension, the value of the exponent of the
power-law plays an essential role in the structural scale invariance;
however, in the case of scale invariant avalanches, the relevance
of the exponent is much less understood. The earthquake dynamics
is the phenomenon that normally comes to people's minds as the example
of scale invariance in the temporal domain. Regardless the value of
the exponent of the power-law distribution, the interpretation of
scale invariance is limited to the absence of characteristic
avalanches, and the existence of many small events and a few very
large ones. Temporal relations between events are sometimes
wrongly added to the interpretation, considering that there is no
correlation between the different avalanches. The logarithmic scale
in which the Gutenberg-Richter law was originally introduced
\cite{Gutenberg-Richter 1956} has also created confusion in the
value of the exponent for the distribution of earthquakes, and
 consequently the implications of this value in the dynamics of
scale invariant avalanches. Further down two examples with different
exponents will clarify that, as in the case of fractal structures,
the exponent of the power-law distribution {\it does} play a central
role in the dynamics of scale invariant avalanches. However, first
we will analyze how to classify scale invariant phenomena, where
the historical use of a logarithmic scale has added some confusion
to the interpretation of scale invariant avalanches.

\subsection{The origin of logarithmic scales}\label{origin_log}

\begin{figure}[b!]
\begin{center}
\vspace{-0.9cm}
\includegraphics{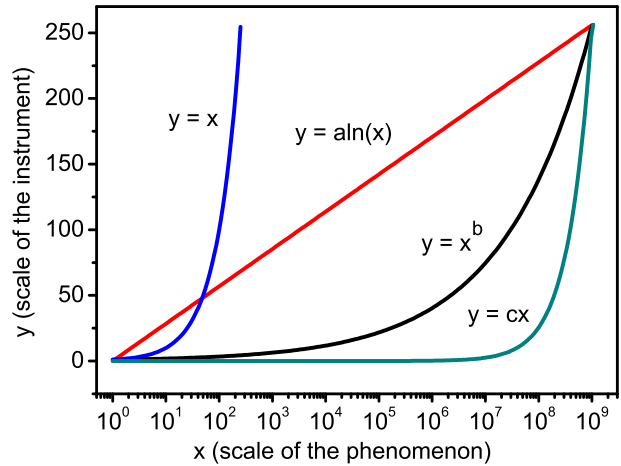}
\end{center}
\vspace{-0.6cm} \caption{\label{fig:scaleTransformations}
Different scale transformation from a phenomenon with $10^9$ levels
of resolution into an instrumental scale of 8 bits of resolution.
 $a=256/ln(10^9)$; $b=ln(256)/ln(10^9)$; $c=256/10^9$.}
\end{figure}

All instruments operate with a finite characteristic resolution,
which need to be consider seriously when analyzing data collected
by the instrument. In a world permeated by scale invariance, how
can one measure a variable presenting values over several orders
of magnitude? The digital music, propelled by a technological
revolution, has focused on increasing the resolution of the
instruments \cite{Peek et al 2009}: 16 bits in a CD, 24 bits in a
DVD, and up to 64 bits in internal processing. Being
able to divide a signal into 32 bits $(4.294...\times 10^9)$, or
even into 64 bits $(1.844...\times 10^{19})$, is extraordinary;
however, nature had to solve this problem with limited
resolution\footnote{The human eye cannot distinguish between 256
grey levels \cite{Russ 2002}.}, and therefore in an ingenious
manner: by the application of scale transformations.
Figure \ref{fig:scaleTransformations} illustrates several
functions transforming a phenomenological scale of $10^9$ levels
of resolution into an instrumental scale of 8 bits of resolution ($256$ levels).

The function $y=x$ saturates the instrumental scale in less than
three decades; while the function $y=cx$, although it covers the whole
range of the phenomena, can not differentiate between values smaller than $10^6$.
In order to find the best function let us introduce a generic transformation function
 $R(x)$ and its inverse $F(y)$. $R(x)$ gives
the change in resolution due to the transformation and $F(y)$ provides
the absolute error $E_{abs}$ and relative error $E_{rel}$ of the
measurement:

\begin{equation}\label{def1}
E_{abs}\equiv dF(y)/dy,~~~~~~~~~~~E_{rel} \equiv
E_{abs}/F(y).
\end{equation}

Two cases are analyzed: $(i)$ a power-law transformation and
$(ii)$ a logarithmic transformation:
\begin{equation}\label{eq01}
    (i)~~~R(x)=a x^b~~~~~~~~~~~~~~~~~~~~~~~~~~~~~~~~~~~~~~~~~~~~~~~~~~~~~~~~~~~~~~~~~~~~~~~~~~~~~~~~~~~~~~~~~~~~~~~~~~~~~~
\end{equation}
\begin{equation}\label{eq02}
    F(y)=(y/a)^{1/b}~,~~~~~~~~~~
    E_{abs}=(1/b)(1/a)^{\frac{1}{b}}~y^{(\frac{1}{b}-1)}~,~~~~
    E_{rel}=(1/b)~y^{-1}
\end{equation}

\begin{equation}\label{eq03}
    (ii)~~R(x)=a~ln(bx)~~~~~~~~~~~~~~~~~~~~~~~~~~~~~~~~~~~~~~~~~~~~~~~~~~~~~~~~~~~~~~~~~~~~~~~~~~~~~~~~~~~~~~~~~~~~~~
\end{equation}
\begin{equation}\label{eq04}
    F(y)=(1/b)~exp(y/a)~,~~~~~~~~~~
    E_{abs}=(1/ab)~exp(y/a)~,~~~~
    E_{rel}=1/a~~~~
\end{equation}

In the case of the power-law transformation, the absolute error
depends on the value of the exponent. For $b=1$ (linear case) $E_{abs}$ is
constant, for $b>1$ it decreases with the measured value, thus the
larger the value the more accurate the measurement. For $b<1$, the
absolute error increases with the measured value. If a phenomenon
occurs over several orders of magnitude, only the case $b<1$ can
fulfil an instrumental scale with limited resolution. In the three
cases the relative error decreases with the measured value. As a
consequence, in the situation of a fractal structure as the one presented in figure \ref{fig:romanesco}; the
larger the measured field, the larger the number of sublevels
resolved by the measurement. Following this reasoning, if a digital camera is used as the instrument of measurement, as the camera moves apart in order
to capture a larger structure, the number of pixels of the camera have to increase, a situation that is normal and common when one uses a tape measure: in order to measure a larger structure, the tape is enlarged and the number of units of measurement increase; thus the relative error decreases. This effect introduces a scale during the process of
measurement, and allows knowing the size of the structure through the
analysis of the relative error; thus, a power-law
transformation ``breaks" the scale invariance. However, the logarithmic
transformation keeps constant the relative error. By using the same
example, the resolution of the camera does not change when the
camera moves apart, and there are no differences between two images
taken at different scales. In this sense a logarithmic
transformation respects the scale invariance, and this is the main
reason for using this scale transformation in the classification of
scale invariant phenomena.

Another reason is historical. In 1856 the English astronomer Norman
R. Pogson proposed the current form of classification of the stars
in different magnitudes in relation to the logarithm of their
brightness \cite{Pogson 1856}. He based the system on the work of
Ptolemy \cite{Ptolemy 140 A.D.}, who probably based his work on the
writings of the ancient Greek astronomer Hipparchus \cite{Toomer
1978}. In 1860 the experimental psychologist Gustav T. Fechner
proposed a logarithmic relation between the intensity of the
sensation and the stimulus that causes it \cite{Fechener 1860}; so
the thought logarithmic response of the human eye\footnote{More recent
studies have proposed power-law relations between sensations and
stimuli, experimentally proved in a rather narrow range of stimuli
\cite{Stevens 1961}.} was responsible for the logarithmic nature of
the stellar scale. In 1935 Charles F. Richter and Beno Gutenberg
proposed a logarithmic scale to describe the earthquake's strength
\cite{Richter 1935}. The name magnitude for this measurement came
from Richter's childhood interest in Astronomy \cite{Los Angeles
Times 1985}; and the scale matches in some degree the earlier
Mercalli intensity scale \cite{Richter 1958}, which quantified the
effects of an earthquake based on human perception.

\subsection{Classifying scale invariant avalanches}\label{classif_avalanches}

Let us analyze now the two examples with different exponents.
Two variables characterize the dynamics of power-law distributed
avalanches: the value of the exponent of the power-law and the
cut-off, which limits the maximum size of the events. In the
following, the analysis will be simplified in considering a sharp
cut-off at a value $S_{max}$.

\begin{equation}\label{deff}
 P(s)=As^b
\end{equation}

\begin{figure}[b!]
\vspace{-1cm}
\includegraphics{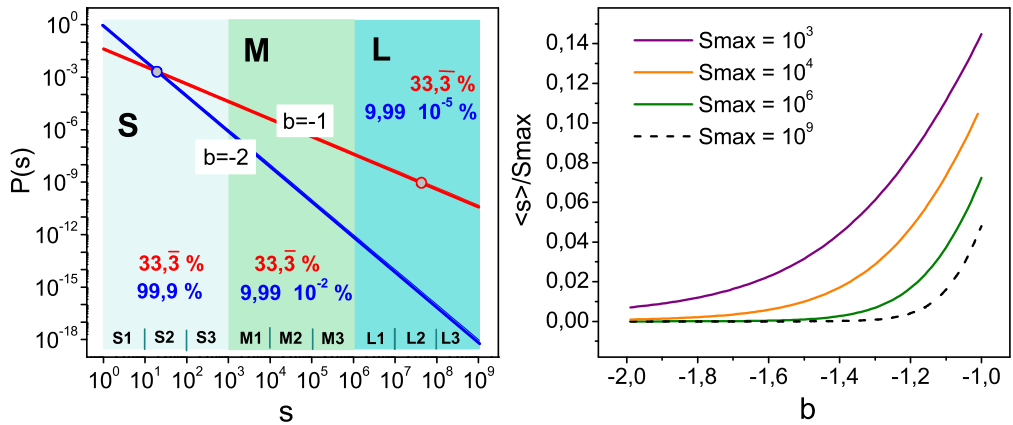}
\vspace{-3.8cm} \caption{\label{fig:2slopes} (a) Power-law
distributions of avalanche sizes for two different exponent values:
$b=-1$ and $b=-2$. The avalanches have been classified as small,
medium and large following logarithmic bins. The percentage of each
type of avalanche for the two different exponents is also displayed.
A circle in each curve represents the mean value of the
avalanche size $\langle s\rangle$. (b) Relation between the mean value of the
avalanche size and the maximum avalanche size $S_{max}$ as a
function of $b$ for different values of
$S_{max}$.}
\end{figure}

\noindent  describes the pdf of the
avalanches, where $1/A=\int_{1}^{S_{max}}s^b ds$ fulfils the
condition of normalization $\int_{1}^{S_{max}}P(s) ds=1$. Figure
\ref{fig:2slopes}a shows the pdfs of two distributions of avalanche
sizes with $S_{max}=10^9$ and exponents $b=-1$ and $b=-2$. The
distributions are represented in a log-log plot, and the avalanches
are classified considering a logarithmic resolution: a ``magnitude" $m$ of the avalanches is defined as the
logarithm of the avalanche size $[m=log(s)]$; and the graph is
divided into $n$ equally spaced zones of $m$. For $n=3$, avalanches
smaller than $m=3$ are considered small $(S)$, those lying between
$m=3$ and $m=6$ are medium $(M)$, and those greater than $m=6$ are
large $(L)$.

\subsubsection{Integrated probability}\label{int_prob}

The main confusion related to the logarithmic scale is a consequence of the fact that during the measurement, an integration has been already performed, which is well described through the integrated probability: $P_{Int}(s,k)=\int_{s}^{ks}As'^b ds$
calculates the probability of having an avalanche with size in
the interval between $s$ and $ks$. Due to the properties of the
integral, the integrated probability is also a power-law with an
exponent $b+1$.

\begin{equation}\label{eq05}
P_{Int}(s,k)=\frac{ln(ks)-ln(s)}{ln(S_{max})}=\frac{ln(k)}{ln(S_{max})}~,~~~~~~~~~~b=-1
\end{equation}
\begin{equation}\label{eq06}
P_{Int}(s,k)=\frac{(k^{b+1}-1)s^{b+1}}{(S_{max}^{b+1}-1)}\simeq
(1-k^{b+1})s^{b+1}~,~~~~~~~~~~-2\leq b<-1
\end{equation}

The calculations performed with a value of $k=10^3$ give the
probabilities of the S, M and L avalanches shown in the graph of
figure \ref{fig:2slopes}a. For $b=-1$ the integrated probability is
constant and equal to $1/n$, so the three types of avalanches have
the same probability equal to $1/3$. In the same manner, considering $k=10$, the graph can be divided into decades: 9 zones equispaced in
$m$ that can be denominated as S1, S2, S3, M1, M2, M3, and L1, L2,
L3 (shown in the graph), all of them with equal probability $1/9$.
This situation, which evidently results from the logarithmic scale of measurement, is very far from the common interpretation of many
small events and only a few very large ones; and in order to
illustrate it more clearly, let us imagine the scenario of the
distribution of earthquakes with an exponent equal to $-1$: the
consideration that one earthquake is happening every second gives in
average one earthquake of magnitude between 2 and 3 (S3 in our
scale) every $9^{th}$ second, but also a catastrophic quake of
magnitude between 8 and 9 (L3 in our scale) every $9^{th}$ second.
Fortunately, many real phenomena with catastrophic consequences have
smaller exponents in their pdfs.

For $b=-2$, $P_{Int}(s,k)=(1-k^{-1})/s$. Every decade the
probability decreases by a factor 10. As a consequence, the
probabilities of having a small avalanche is $0.999$; $9.99 \times
10^{-5}$ for a medium size avalanche; and only $9.99 \times 10^{-7}$
for a large event (figure \ref{fig:2slopes}a). Again, if we imagine
the scenario of the distribution of earthquakes with an exponent
equal to $-2$: the consideration that one earthquake is happening
every second gives on average one minor earthquake of magnitude between 2
and 3 (S3 in our scale) every $111$ seconds, one moderate of magnitude between 5
and 6 (M3 in our scale) every $1.1\bar{1} \times 10^5$ seconds (30,8 hours), and a catastrophic
quake of magnitude between 8 and 9 (L3 in our scale) every
$1.1\bar{1} \times 10^8$ seconds (3,5 years).

\subsubsection{Mean value of avalanche size}\label{mean value_avalanches}

Another relevant quantity signaling the key role of the exponent of
the power-law, corresponds to the mean value of the size
distribution of the avalanches $\langle s \rangle=\int_{1}^{S_{max}}sP(s)ds$.

\begin{equation}\label{avasize01}
\langle s \rangle=\frac{S_{max}-1}{ln(S_{max})}~,~~~~~~~~~~~~~~b=-1
\end{equation}
\begin{equation}\label{avasize02}
\langle s \rangle=\frac{(b+1)}{(b+2)}\frac{S_{max}^{(b+2)}-1}{S_{max}^{(b+1)}-1}~,~~~~~~~~~~~-2<b<-1
\end{equation}
\begin{equation}\label{avasize03}
\langle s \rangle=\frac{ln(S_{max})}{1-S_{max}^{-1}}~,~~~~~~~~~~~~~~b=-2
\end{equation}

The mean value of the avalanche size is related to both the response
of the system to a perturbation, and the energy balance in the
dynamics. In the figure \ref{fig:2slopes}a, where $S_{max}=10^9$,
the values of $\langle s \rangle$ correspond to $4.8 \times 10^7$ and $20.7$ for
$b=-1$ and $b=-2$ respectively. These values are represented by a
circle in each curve.

The value of $\langle s \rangle$ corresponds to the average response of the system to
a perturbation, under the consideration that small perturbations can
provoke the overcoming of local thresholds and thus the triggering
of avalanches. In average, the system is delivering an avalanche of
size $\langle s \rangle$; so in terms of avalanche production, this is equivalent
to generate an avalanche of size $\langle s \rangle$ in every event of the
dynamics.  In the particular case of $\langle s \rangle$ proportional to the
system size, the situation can be interpreted as critical: in average
 a perturbation provokes a response proportional to the system size. However, the fact that the dimension of the avalanche is smaller than the dimension of the system, adds some complexity to the analysis of the criticality through the avalanche size distribution, which will be discussed in section~\ref{dissip and struct}

\subsubsection{Energy balance in slowly driven systems}\label{energy balance_avalanches}

As mentioned in the
introduction, avalanches are defined as sudden liberation of energy
that has been accumulated very slowly. This indicates that the
energy is injected in small portions, and that there is a separation between the drive of the system
(slow) and the avalanche duration (rapid). At every single time interval, it is possible to define an injected energy, an avalanche of a particular size, and a dissipated energy.
If the system is in a
stationary state, the average energy injected to the system in every
time interval has to be equal to the average dissipated energy.
Consequently, the average dissipated energy has to be small,

\begin{equation}\label{eq10}
\langle E_{injected} \rangle = \langle E_{dissip} \rangle.
\end{equation}

Many of the models dealing with scale invariant avalanches are non-dissipative in
the bulk, and the energy is liberated through the boundaries of the
system \cite{BTW 1987}. However, they still refer as avalanches the local processes
related to rearrangements in the bulk of the system, with no energy
cost. As the avalanche production is not directly related to the
dissipation of energy, these systems can have a large value of $\langle s \rangle$
and still present a small average value of the dissipated energy $\langle E_{dissip} \rangle$.

However, the vast majority of real phenomena are dissipative.
Considering that
\begin{equation}\label{eq11a}
\langle E_{injected}\rangle \sim S_{min} << S_{max},~~~~and~~~~
 \langle E_{dissip} \rangle = \alpha \langle s \rangle,
\end{equation}

\noindent where $\alpha$ is a dissipation coefficient,

\begin{equation}\label{eq11}
\alpha \langle s \rangle <<S_{max}.
\end{equation}

The larger the avalanche size, the larger the dissipation; and as a
consequence, large values of $\langle s \rangle$ are forbidden for dissipative
slowly driven system. Figure \ref{fig:2slopes}b shows the
relation $\langle s \rangle / S_{max}$ for different $b$ values, indicating that
large values of $b$ (close to $-1$) are forbidden for dissipative
slowly driven system.

The previous analysis did not consider the existence of
avalanches of size zero (let us call them {\it zero} avalanches),
where the addition of energy to the system provokes no response in
terms of avalanches ($s<S_{min}$). In order to compensate the
energy lost for an average {\it non-zero}  avalanche $\langle s \rangle$, the system needs a number  $n_{S0}$ of {\it
zero} avalanches proportional to $\alpha \langle s \rangle / S_{min}$.
For small $b$ values, and thus small $\alpha \langle s \rangle$, $n_{S0}$ can be the consequence of a lack of resolution in the measurement.

\subsubsection{The distribution of earthquakes}\label{earthquakes}

The Gutenberg-Richter law \cite{Gutenberg-Richter 1956} is the best known example of scale invariant avalanches. However, many studies from the Statistical Physics point of view have not considered the fact that the original distribution was measured in a logarithmic scale, which provokes a change equal -1 in the measured exponent \cite{Bak and Tang 1989, Sethna 2009}. Other reports attribute this change to a cumulative manner in the definition of the probability \cite{Bak et al 2002}, which is mathematically correct, but it is not the right interpretation; and it brings confusion to the Geological community \cite{Main 1999}.

\begin{figure}[h!]
\begin{center}
\vspace{-0.5cm}
\includegraphics{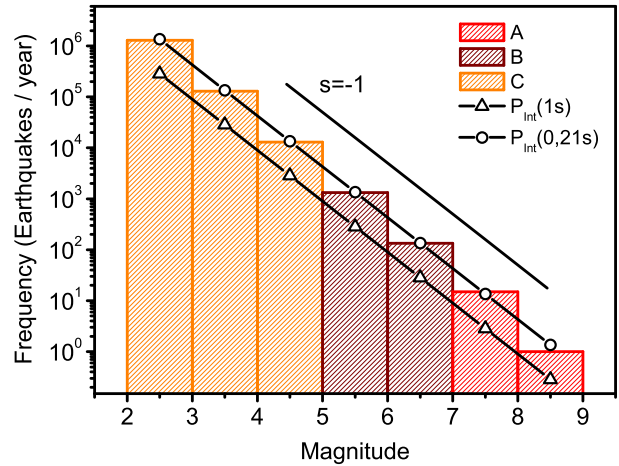}
\end{center}
\vspace{-1cm} \caption{\label{fig:quakes}
Frequency of earthquake occurrence worldwide. Data from the U.S. Geological Survey \cite{USGS statistics}: (A) Based on observations since 1900. (B) Based on observations since 1990. (C) Estimated. A solid line with slope $s=-1$ serves a a guide. The solutions of the Integrated probability for $b=-2$, considering that one earthquake occurs every second, and every 0.21 seconds, are also displayed.}
\end{figure}

Figure \ref{fig:quakes} displays the frequency of earthquake occurrence worldwide. The fact that the results are presented in intervals of magnitude indicates that the distribution corresponds to the integrated probability $P_{Int}(M_L)$, where $M_L$ is the local magnitude. The graph shows that $log[P_{int}(M_L)] \sim -M_L$. By substituting $M_L$ into the definitions

\begin{equation}\label{eq12}
M_L \equiv log(A)+const
\end{equation}
\begin{equation}\label{eq13}
M_L \equiv 2/3~log(E)+const
\end{equation}

\noindent where $A$ (wave amplitude) is the maximum excursion of the Wood-Anderson seismograph, according to the original work of Richter \cite{Richter 1935}, and $E$ the energy released by a quake, one gets $P_{Int}(A_L, 10) \sim A^{-1}$ and $P_{Int}(E_L, 10) \sim E^{-2/3}$. By adding $-1$ to the values of the exponents in the integrated probability, as discussed earlier (section~\ref{int_prob}), it is possible to obtain the distributions of earthquakes in terms of $A$ and $E$:

\begin{equation}\label{eq14}
P(A) \sim A^{-2},~~~~~~~~~~~~~~P(E) \sim E^{-5/3}
\end{equation}

In order to illustrate the integrated probability for the distributions of figure \ref{fig:2slopes} (section~\ref{int_prob}), the consideration of one earthquake happening every second was used. The results of this assumption for the case $b=-2$ are plotted in figure \ref{fig:quakes}, and they are a little lower than the real values. However, the same analysis under the consideration of one earthquake occurring every $0.21$ seconds fits quite well the real data (figure \ref{fig:quakes}).

\section{Avalanches in Critical Phenomena}\label{ava_critical phenom}

The original motivation of this section was to study the properties of the scale invariant avalanches in well established critical systems, in order to compare them with the SOC avalanches. As discussed in the introduction, the SOC borrowed the concept of critical
point of equilibrium phase transitions in order to describe their
uncorrelated power-law distributed avalanches. The term {\it
critical} in the avalanche context has been presented through the
fact that at any moment a minor perturbation can trigger a response
(avalanche) of any size and duration, a behavior that is linked to a
divergence of the correlation length in the original numerical model
of SOC: the BTW model \cite{BTW 1987}. Many dissipative
phenomena involving avalanches distributed according to power-laws
have been treated as critical systems \cite{Jensen 1998, Bak 1997}; however,
recent studies have shown different systems evolving through power-law
distributed avalanches in a non-critical behavior \cite{Ramos et al 2009, Ramos 2010, Pruessner and Peters 2006}, which has motivated the analysis of the avalanche dynamics in a
well established critical scenario: a second order phase transition.

In Physics, the classical scenario of critical phenomena takes place during a second order phase transition \cite{Stanley 1987}. The text-book example is the transition where a permanent magnet loses its magnetism:  its magnetic properties cease when the temperature is increased above a certain critical temperature $Tc$. Below this temperature, a majority of spins point in the same direction, creating a magnetic field. Large fluctuations in spins do not occur at low temperatures, thus the system will remain unchanged. Above the critical temperature the spins' directions are random and change direction randomly, frequently and individually. The system is already disordered, therefore, no large-scale changes will happen; there is no overall magnetic field. However, at the critical temperature itself, large fluctuations occur and different snapshots of the system show different patterns - but all of the patterns will be statistically similar, in that clusters of aligned spins are surrounded by areas with spins oriented in the opposite direction. The clusters are of all sizes and their distribution follows a power-law \cite{Clusel et al 2004}. Four characteristic of this critical state will be used in our analysis along this chapter:

a) Divergence of the correlation length ($\xi$): The temporal average of the spatial autocorrelation function

\begin{equation}
\label{spatial corr_funct}
\langle C_As(d,t) \rangle _t= \Bigg \langle \frac{\sum f(x,y)f(x',y')- \langle f(x,y) \rangle ^2}{\sum (f(x,y)- \langle f(x,y) \rangle)^2} \Bigg \rangle _t
\end{equation}

\noindent where $f(x,y)$ represents the structure of the system (in two dimensions) and $d$ corresponds to the distance between $(x,y)$ and $(x',y')$, can generally be fitted ($\forall T\neq Tc$) as an exponential decay in the form

\begin{equation}
\label{corr length}
\langle C_As(d,t) \rangle _t \sim exp(-d/\xi).
\end{equation}

At the critical point, the correlation length $\xi$ is proportional to the linear size of the system $L$ (diverging in an infinite system). The temporal average is necessary because the calculus is performed in a snapshot of the dynamics (a microstate), and any physical measure implies an average over many different microstates, which is equivalent to a temporal average if the system is ergodic \cite{Newman and Barkema 1999}.

b) Divergence of the correlation time ($\tau$): The temporal autocorrelation function

\begin{equation}
\label{time corr_function}
C_At(t) = \frac{\sum f(t_i)f(t_i+t)- \langle f(t_i) \rangle ^2}{\sum (f(t_i)- \langle f(t_i) \rangle) ^2}
\end{equation}

\noindent can be fitted as an exponential decay in the form

\begin{equation}
\label{corr time}
C_At(t) \sim exp(-t/\tau).
\end{equation}

Far from the critical point, the correlation time $\tau$ is small, so the system will quickly recover from a perturbation. At the critical point, $\tau$ diverges due to the fact that the system hesitates between the two states, and perturbations can move the system away from its equilibrium state during long periods of time. As a result, the dynamics turns slow, a phenomenon which is known as {\it critical slowing down} (CSD) \cite{Newman and Barkema 1999}.

c) Both $\xi$ and $\tau$ present power-law dependencies with the reduced temperature $T_r=(T-Tc)/Tc$ in the way $\xi = | T_r |^{- \nu}$ and $\tau = | T_r |^{- z \nu}$; and thus they relate to each other through $\tau = \xi^{z}$. $\nu$ is called a {\it critical exponent} and is an attribute of the Ising model. Phenomena with the same critical exponents belong to the same universality class. The exponent $z$ is often called the dynamic exponent. It gives a way to quantify the CSD and it depends on the algorithm, i.e., it depends on the type of dynamics \cite{Newman and Barkema 1999}.

d) As the size of the system increases, the transition between the two states becomes sharper, and it is infinitely sharp in an infinite system \cite{Newman and Barkema 1999}.

\subsection{Avalanches $\neq$ fluctuations}\label{ava_fluct}

\subsubsection{Simulation: the Ising model}\label{ising model}

In this subsection, the dynamics of avalanches is studied in a well established critical system: the Ising model, which is certainly the most thoroughly
researched model in the whole of statistical physics. It is a model
of a magnet, and consists of a lattice where every site represents a
spin of unit magnitude taking two values $\pm 1$ as an indication of
the only two possible directions to point: ``up" or ``down". The
spins interact with their nearest neighbors and the magnetization
$M$ is the sum over all the spins. For two or more dimensions the
system shows a second order phase transition at a critical
temperature $Tc$ ($Tc=2.269$ in two dimensions) from a
ferromagnetic to a paramagnetic state when the temperature is
increased, and in two dimensions the model is analytically solved
\cite{Onsager 1944}. The behavior of the average fluctuations of
the magnetization is well known, and it defines the magnetic
susceptibility, as well as its relation with the correlation
function. The fluctuations of the magnetization ($M- \langle M \rangle$) have also
been studied \cite{Stauffer 2000}, and their pdf is reported to be
universal \cite{Bramwell et al 2000}. It presents an exponential
tail on one side, and a rapid falloff on the other side. However, in
the scenario of SOC systems, instead of analyzing the fluctuations
of the magnitudes, the standard is to define avalanches,
corresponding to relative differences between consecutive states. Therefore, the definition of
avalanches presented in the introduction, which is relative to slowly
driven system, has been extended to {\it differences between equispaced values in the time series of the measured variable}. In the case of the Ising model, they corresponds to jumps in the magnetization between two consecutive
microstates. Small simulations, both in size and in running time,
will be sufficient to illustrate the dynamics of avalanches in
this critical scenario.

\begin{figure}[b!]
\begin{center}
\vspace{-0.5cm}
\includegraphics{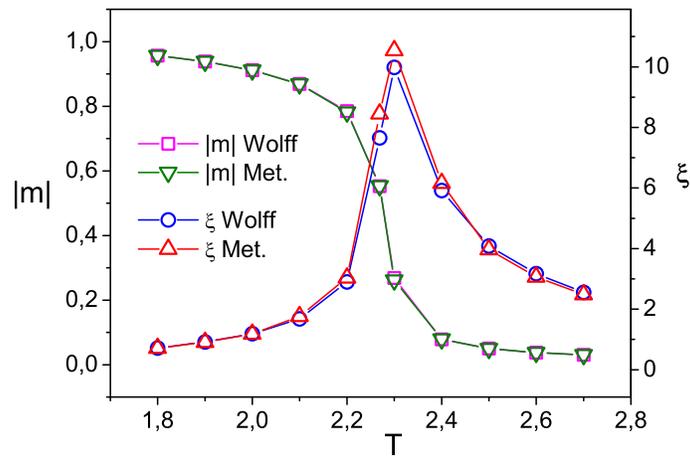}
\end{center}
\vspace{-1cm} \caption{\label{fig:corr_ising}
Average of the
absolute value of the magnetization per spin and correlation length
for both Metropolis and Wolff algorithms at different temperatures.}
\end{figure}

The simulations take place in a $128 \times 128$ lattice with
periodic boundary conditions, and compute $10^{6}$ Monte Carlo steps
(MCS) after $10^{5}$ thermalization steps. Both Metropolis
\cite{Metropolis et al 1953} and Wolff \cite{Wolff 1989}
algorithms are implemented considering no external magnetic field.
Thus the energy of the system reads as $E=-J \sum_{\langle i,j \rangle }^N s_i s_j$
where $J=1$ is a coupling constant and $\langle i,j \rangle$ indicates that the
sum is over nearest neighbors only. In the Metropolis algorithm one
MCS consists of $N=(128)^2$ events where one random spin is
selected, and flipped ($s_i=-s_i$) if $exp(-\Delta E/KT)$ is larger
or equal to a random number between 0 and 1. $\Delta E$ is the
change in energy due to the flip of the spin, and K is considered
equal to 1, so the temperature T is presented as an adimensional
magnitude. In the Wolff algorithm, one MCS consists of building up a
cluster of flipped spins. Starting from flipping one spin at a
random position, its neighbors will become part of the growing
cluster if $exp(-2J/KT)$ is smaller than a random number between 0
and 1.

\begin{figure}[b!]
\begin{center}
\vspace{-1.2cm}
\includegraphics{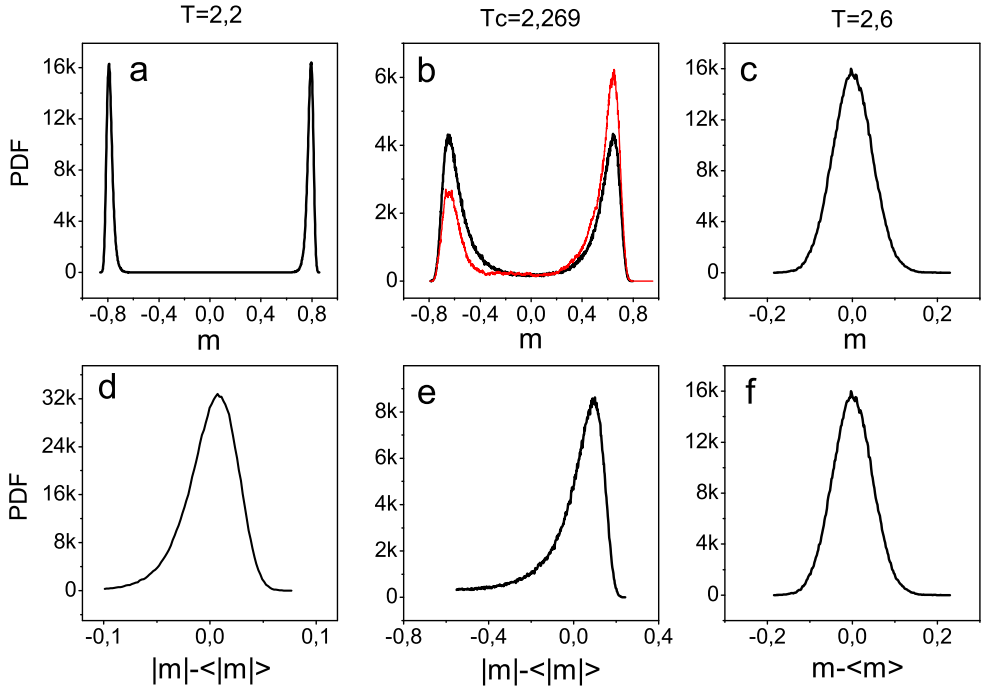}
\end{center}
\vspace{-1.5cm} \caption{\label{fig:fluct_ising}
(a-c) Distribution of the magnetization for the Wolff algorithm at different temperatures. At the critical point, the distribution for the Metropolis algorithm is also displayed (b), presenting an asymmetric behavior. (d, e) Distribution of the
fluctuations of the absolute value of the magnetization for the Wolf algorithm at different temperatures. (f) Distribution of the fluctuations of the magnetization for the Wolff algorithm at T=2.6.}
\end{figure}

The average of the absolute value of the magnetization per spin $m$,
and the value of the correlation length $\xi$ are shown in
figure~\ref{fig:corr_ising}. The correlation length has been extracted
following the eqs.~\ref{spatial corr_funct} and~\ref{corr length}:

\begin{equation}
\label{corr ising}
\langle C(r) \rangle _t = \Bigg \langle \frac{\sum_{d(i,j)=r}^N ( s_i s_j)-m^2}{\sum_{d(i,j)=r}^N (s_i |s_j|-m)^2} \Bigg \rangle _t
\end{equation}
\begin{equation}
\label{corr ising}
\langle C(r) \rangle _t \sim exp(-r/\xi).
\end{equation}

The correlation function has been calculated by using periodic
boundary conditions, through an average of values taken every
thousand MCS, and the sum is over those sites that are separated
from each other by a distance equal to $r$. Both algorithms give
very similar results on the averages of the physical magnitudes,
with a peak in the correlation length coinciding with the phase
transition.

The behavior of the fluctuations for the Wolf algorithm is
presented in figure~\ref{fig:fluct_ising}. For very low temperatures
two symmetric Gaussian distributions (GDs) in the pdf of the magnetization indicate
the two symmetric ordered states in the system (not shown in the graph). As the temperature
increases, the two GDs approach each other (figure~\ref{fig:fluct_ising}a) and a small asymmetry starts
to be noticed at the low frequencies in the pdf of the fluctuations of the
absolute value of the magnetization (figure~\ref{fig:fluct_ising}d).
At the critical point, the two GDs start to merge forming the universal Gumbel distribution
reporter by Bramwell {\it et. al.} in the pdf of the fluctuations of the
absolute value of the magnetization  \cite{Bramwell et al 2000, Bramwell et al 1998} (figure~\ref{fig:fluct_ising}e). This Gumbel distribution of the fluctuations has been used by different experiments as an indication of the {\it criticality} of the system \cite{Joubaud et al 2008, Planet et al 2009}. For high temperatures we can consider that the two GDs have perfectly merged, and there is a Gaussian behavior of the fluctuation of the magnetization (figure~\ref{fig:fluct_ising}f). In the case of the Metropolis algorithm only one side of the graph ($a$) will be explored by the system in a finite time (either positive or negative magnetization). The asymmetry displayed by the Metropolis algorithm in the graph ($b$) is a consequence of its slow dynamics, as it will be discussed further down. $10^{6}$ MCS are not enough for
the system to spend on average the same ``time" in symmetrical areas
of the phase space (with $10^{7}$ MCS both algorithms have given the
same result). The other graphs ($c-f$) are the same for both algorithms; consequently, they give the same results concerning
the fluctuations of the absolute value of the magnetization.

\begin{figure}[h!]
\begin{center}
\vspace{-0.8cm}
\includegraphics{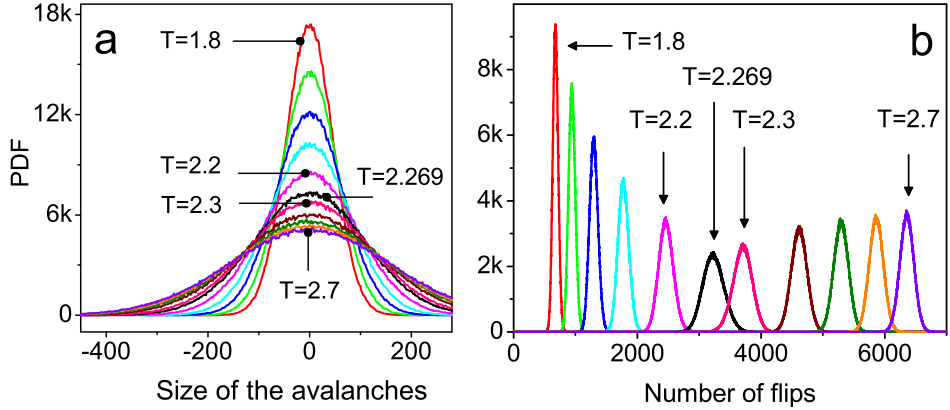}
\end{center}
\vspace{-4cm}\caption{\label{fig:metropolis} a) Distribution of
the jumps in the magnetization (avalanches) for different temperatures (T) with
the Metropolis algorithm. The curves correspond to T values between
1.8 and 2.7 with spacings of 0.1, and also the critical point:
$T_{c}=2.269$. b) Distributions of flips
for different temperatures. }
\end{figure}

The distributions of the jumps of $M$ (avalanches) and the
distributions of flips for the Metropolis algorithm are displayed in
figure~\ref{fig:metropolis}. The avalanches follow a Gaussian for every measured
T, and the distributions widen as T increases (figure~\ref{fig:metropolis}a). The distributions of
flips (figure~\ref{fig:metropolis}b) also follow Gaussian
distributions, with their mean values increasing with T and standard
deviations having a maximum at the critical value $Tc=2.269$.

\begin{figure}[t!]
\begin{center}
\vspace{-0.8cm}
\includegraphics{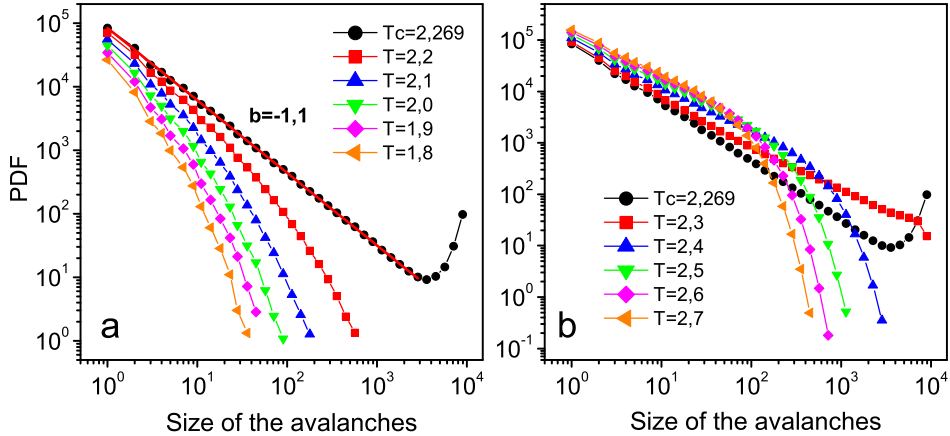}
\end{center}
\vspace{-4cm}\caption{\label{fig:wolff} Distributions of the jumps
of the magnetization (avalanches) for the Wolff
algorithm at different temperatures. }
\end{figure}

The jumps of $M$ in the Wolff algorithm (avalanches) are displayed in
figure~\ref{fig:wolff}. The way that clusters build-up makes the
absolute value of the jumps of $M$ equal to the number of flipped
spins, so only the distributions of avalanches are presented. As
T decreases from the critical point the distributions display an
increasing number of events that involve the whole system, which
is an artefact. They have been removed from the graph and will not be taken
into account in the analysis. At the critical temperature the
avalanches follow a power-law distribution with an exponent
$b=-1.1$. The distributions deviate from power-laws
as the temperature moves away from $Tc$.

The analysis of the simulations has demonstrated that power-law
distributed avalanches are not a necessary condition in order to
classify a system as {\it critical}, but different kinds of
distributions can rule the avalanche behavior of an equilibrium
system at the critical point. The Metropolis dynamics at $Tc$ is
ruled by avalanches whose sizes are distributed following a Gaussian with
standard deviation much smaller than the system size. It is a local
dynamics happening on (and slowly re-shaping) a globally correlated
landscape. Small jumps will move the system slowly around the phase
space, so in a real situation a fast enough measurement can get the
system ``trapped" on a fluctuation, even far from the equilibrium
value (calculated after a long enough averaging). This slow dynamics is
directly linked to the CSD, situation which has been artificially eliminated in the Wolff algorithm. The Wolff dynamics
at $Tc$ is the one that people working with SOC models are
accustomed to: avalanches that follow the landscape in an
``invasion" mode, cascading through the system and resembling the
properties of the globally correlated landscape. This direct relation between structure and avalanches is the key that allows analyzing the critical properties of a system from the characteristics of its scale invariant distribution of avalanches.

Two questions emanate from this analysis. The first: Is there an ``algorithm" (a dynamics) chosen by nature? And the second: Is there CSD in real critical systems evolving through scale invariant avalanches? Let us discuss some experimental results.

\subsubsection{Experiments: from the micro to the macro-world}\label{experiments_fluct_avalanches}

The Wolff algorithm is much faster than the Metropolis one around $Tc$. This was a
big achievement in 1989, where the computing capabilities where rather limited. However,
it is well known that Metropolis represents better the dynamics of the second order phase transition around the Curie temperature \cite{Dunlavy and Venus 2005}. The reason for that is the fact that many other less stronger interactions have not been considered in the analysis. They are included in something denominated {\it thermal bath} which tries to equilibrate the dynamics. Recent experiments have focused on the non-gaussian (Gumbel) distribution of fluctuations close to the Fr\'eedericksz transition, a second order phase transition in a liquid crystal \cite{Joubaud et al 2008}, and they have also confirmed the Gaussian character of the avalanche distributions close to the critical point \cite{Ciliberto 2009}. However, if the system is kept at a low temperature (below $Tc$) and an external magnetic field is applied, the spins try to align themselves with the external field, and a reorganization of the magnetics domains takes place. This reorganization is not a smooth process, but is composed of small bursts or avalanches distributed following a power-law; a phenomenon which has been widely studied from the avalanche context \cite{Sethna et al 1997, Sethna et al 2001, Zapperi et al 1998} and is known as the Barkhausen effect \cite{Barkhausen 1919}.

\begin{figure}[t!]
\begin{center}
\vspace{-0.6cm}
\includegraphics{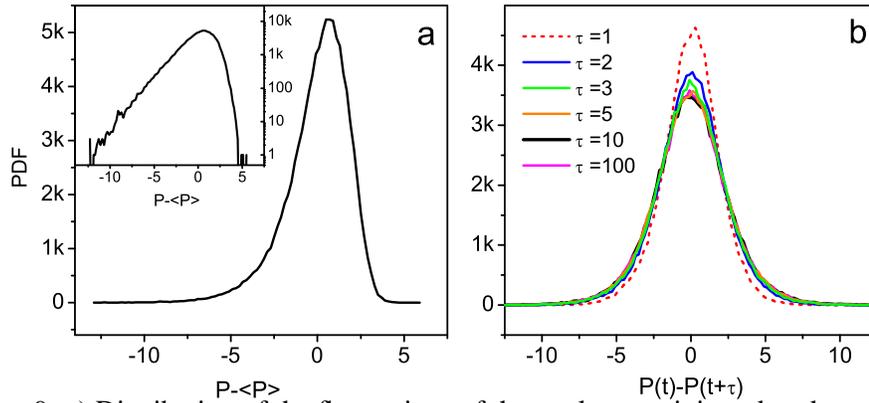}
\end{center}
\vspace{-4.5cm}\caption{\label{fig:pinton} a) Distribution of
the fluctuations of the total power injected to the motors of
the rotatory disks in a turbulent flow experiment. Inset: The same function in log-lin axes,
in order to enhance the asymmetry of the Gumbel. b) Distributions
of avalanches for different $\tau$ values.}
\end{figure}

Moving towards the macro-world, a turbulent flow is another phenomenon where non-gaussian fluctuations have been reported \cite{Bramwell et al 1998, Labbe et al 1996}: two coaxial disks counter-rotate at a fixed velocity generating a swirling flow in the gap between them. The power required to keep a constant velocity of the disks (through a feedback loop) is measured and the fluctuations of the power correspond to the variable of analysis. If the experiment is performed inside a cylinder coaxial with the disks but with a diameter much larger than the disks' diameter, the fluctuations of the total power follow a Gaussian distribution. However, if the diameter of the cylinder is only slightly larger than the disks' diameter, the fluctuations of the total power follow a Gumbel distribution.The data presented in \cite{Bramwell et al 1998} have been reanalyzed for this chapter (courtesy of J.-F. Pinton): avalanches have been defined as relative differences between two points separated a time interval $\tau$ in the time series of total power. While the fluctuations of the total power display a Gumbel distribution (figure~\ref{fig:pinton}a), interpreted as a signature of a critical scenario; the avalanches follow a Gaussian distribution for all the different $\tau$ values (figure~\ref{fig:pinton}b). A remarkable difference in relation to the Ising model is the fact that the standard deviations of the Gaussian distributions are comparable to the width of the Gumbel. By following the same reasoning used in the Ising model, the Gumbel can be explained as the merging of the two Gaussian distributions of the fluctuations of the individual disks due to the confinement, which reduces the number of degrees of freedom in the system. However, the explanation of the Gaussian distribution in the avalanches and its relation to the properties of the flow are questions that surpass the scope of this chapter.

\begin{figure}[t!]
\begin{center}
\vspace{-0.25cm}
\includegraphics{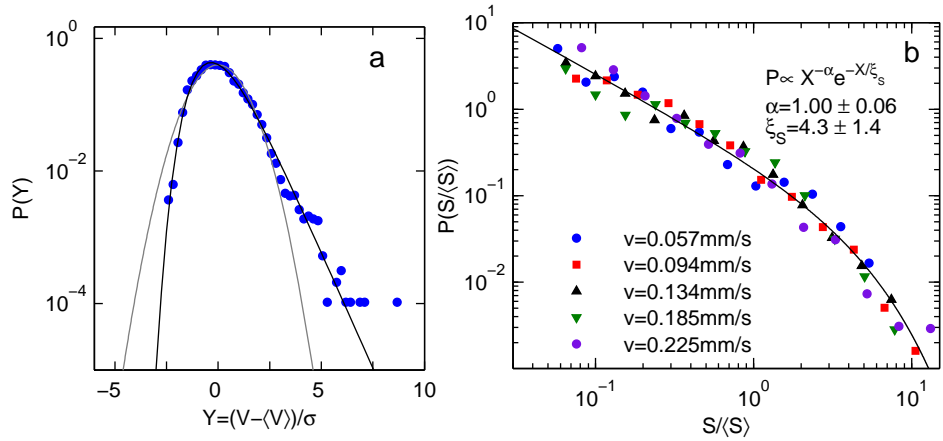}
\end{center}
\vspace{-1.2cm}\caption{\label{fig:stef} a) Distribution of
the fluctuations of the front velocity for a low injection rate equal to $v=0.057 mm/s$ in an imbibition experiment.
(b) Distributions of avalanches of the front velocity, normalized to the temporal average of the front velocity, for different injection rates (Courtesy of R. Planet and S. Santucci).}
\end{figure}

Recently an experiment has been reported where both avalanches and fluctuations have been measured in an imbibition front \cite{Planet et al 2009}. A viscous liquid is injected into a Helle-Shaw cell where a random distribution of well controlled patches guaranties a specific level of disorder. The competition between viscous and capillary forces creates a jerky dynamics where the global velocity of the front displays avalanches that follow a scale invariant distribution fitted to an exponent $-1$ (figure~\ref{fig:stef}b). The average of the avalanche size increases with the decrease of the injection rate ($v \rightarrow 0$) and as a consequence, the cutoff of the power-law moves to higher values of avalanche size, approaching a pure power-law in the limit ($v=0$). The distribution of fluctuations also change from a Gaussian to a Gumbel as $v \rightarrow 0$ (figure~\ref{fig:stef}a), indicating the critical properties of the system around $v=0$. The authors have been able to relate the asymmetry of the Gumbel to the reduction of the degrees of freedom in the system as the velocity of the dynamics is reduced.

The main message of this part consists in revealing that power-law
distributed avalanches are not a necessary condition in order to
classify a system as {\it critical}, but different kinds of
distributions can rule the avalanche behavior of an equilibrium
system at the critical point. The question about the dynamics chosen by nature has been answered in the way that different dynamics have been observed in diverse phenomena; nevertheless, a general formalism capable of predicting the kind of dynamics for a particular phenomenon is still lacking. Concerning the question about CSD, although in both the Barkhausen and the imbibition experiments the critical point of the pinning-depinning transition is reached at a limit ($v \rightarrow 0$), the fact that the velocity of the dynamics is externally imposed leaves some doubts about the existence of CSD in real critical systems evolving through scale invariant avalanches. A stronger argument in favor of its existence comes from percolation, which shares many different points with a second order phase transition (including CSD) at the percolation threshold, and where the avalanches are power-law distributed \cite{Stauffer and Aharony 1994}. Another thing to retain from this part is the value of the exponent of the power-law displayed by a well established critical system in two dimensions.

\section{Criticality in Scale Invariant Avalanches}\label{critic_avalanches}

\subsection{Models without spatial structure}

After the introduction in 1987 of the SOC \cite{BTW 1987} as a new critical phenomenon occurring in a class of dissipative coupled systems triggered by temporal fluctuations, many different approaches have been used in order to describe its dynamics. The first one corresponded to the {\it critical branching process} \cite{Alastrom 1988}.

\begin{figure}[h!]
\begin{center}
\vspace{-2.0cm}
\includegraphics{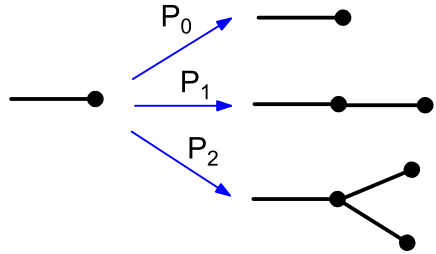}
\end{center}
\vspace{-4cm} \caption{\label{fig:branching}
Branching process with $n_{max}=2$.}
\end{figure}

The schema in figure~\ref{fig:branching} represents a branching process where each branch has a probability $P_n$ of having n subbranches,  $n\in [0, n_{max}]$. The different probabilities can be calculated by the following equations:

\begin{equation}
\label{branching1}
\sum_{0}^{n_{max}} nP_n = 1+G
\end{equation}
\begin{equation}
\label{branching2}
\sum_{0}^{n_{max}} P_n = 1
\end{equation}

\noindent where $G$ corresponds to the average growth. {\it Criticality} is reached in the situation where the process barely ``survives" \cite{Alastrom 1988}, which corresponds to $G=0$; i.e., it is the minimum probability able to develop branches proportional to the system size $L$; where $L$ is to the length where the process is artificially stopped. For $n_{max}=2$, $P_0=P_2=(1-P_1)/2$. If avalanches are defined as the number of generated branches, and the process is repeated a large number of times, the avalanche size distribution corresponds to a power-law with an exponent equal $-3/2$ and an exponential cut-off in the form

\begin{equation}
\label{branching3}
P(s) \sim s^{-3/2}exp(-s/\lambda)
\end{equation}

\noindent where $\lambda \sim L$ \cite{Harris 1963}. This result is in perfect agreement with the {\it mean-field theory} \cite{Vespignani and Zapperi 1997}, with {\it percolation} in a Bethe lattice \cite{Stauffer and Aharony 1994}; and also with recent works using {\it functional renormalization group} \cite{Le Doussal Wiese 2009}. In the first two approaches there is no spatial structure. However, the same result appears if the processes take place in a real lattice, but only above a critical dimension $d_c$. The higher the dimension, the less the probability for a branch to form a loop; and this absence of loops is a necessary conditions that makes the calculation possible. Unfortunately, $d_c=4$ for the branching process \cite{Obukhov 1988, Diaz-Guilera 1994}, and for the functional renormalization group \cite{Le Doussal Wiese 2009}. For percolation $d_c=6$ \cite{Stauffer and Aharony 1994}. However, in real situations of two and three dimensions the results are different, and we will find values smaller than -3/2 for the critical exponents.

\subsubsection{The role of dissipation and the structure of the avalanches}\label{dissip and struct}

Although the solution displayed by eq.~\ref{branching3} works for spatial dimensions beyond the real world, it is very useful both for setting the lower limit to the critical exponents to -3/2, and for understanding the main concepts through discussion on an analytic base. By setting a negative value to $G$ in eq.~\ref{branching1} it is possible to simulate the effect of the dissipation during the branching process: At every branching occasion, the probability is lower than the critical one; and as a result the average length of the branches decreases. The solution of the avalanche size distribution has the same shape (eq.~\ref{branching3}), but with the difference that $\lambda$ decreases with the dissipation, results that have also been reported in a Bethe lattice \cite{Lauritsen et al 1996}. Dissipation reduces the size of the avalanches, and as a consequence the linear size of the mean avalanche is not proportional to the linear size of the system. $\lambda$ is normally considered as the correlation length $\xi$, with the implication that the system is critical only in the conservative case.

There is a general belief that a dynamic of power-law distributed avalanches is a signature of a critical scenario (independently of the value of the exponent), and that the existence of a cut-off is the indication of the loose of critical properties. This consideration is based on the fact that, regardless the value of the exponent, eventually an avalanche is reaching the system size ($s_t\sim L$). However, in the analysis of the correlation length in section~\ref{ava_critical phenom}, the necessity of the temporal average has been presented. A particular avalanche ($s_t$) corresponds to a ``microstate" in the dynamics, and a temporal average is needed in order to get the values of the physical magnitudes. Let us analyze this in detail:

The correlation length $\xi$ has been defined in the eqs.~\ref{spatial corr_funct} and~\ref{corr length}; and the {\it criticality} of a system has been introduced as the divergence of the correlation length ($\xi\sim L$). The analysis of the Wolff algorithm in section~\ref{ising model} has shown a strong relation between structure and avalanche dynamics, suggesting that in a dynamics of scale invariant avalanches, it is possible to measure the {\it criticality} of the system through the average value of the avalanche sizes:
\begin{equation}
\label{dim01}
\langle s \rangle^{1/d_A}\sim L,
\end{equation}
\noindent where $d_A$ corresponds to the fractal dimension of the avalanche in a system of dimension $d_S$, and volume $V\sim L^{d_S} $. From eq.~\ref{avasize02} one gets $\langle s \rangle\sim S_{max}^{b+2}$; therefore,
\begin{equation}
\label{dim02}
S_{max}^{(b_c+2)/d_A}\sim L\sim V^{1/d_S} \geq S_{max}^{1/d_S},
\end{equation}
\begin{equation}
\label{dim021}
(b_c+2)d_S\geq d_A.
\end{equation}

A larger value of the exponent $b$ provokes a larger value of $\langle s \rangle$, and consequently it delivers larger avalanche sizes. By following the same principle of the branching process, where the criticality was defined as the lower probability able to form branches reaching the system size, it seems possible to choose the smaller $b_c$. Consequently,

\begin{equation}
\label{dim03}
(b_c+2)d_S=d_A.
\end{equation}

The relation~\ref{dim03} links the critical exponent $b_c$ to the difference between the fractal dimension of the avalanche and the dimension of the system. If $d_A=d_S$, the critical exponent is equal to $-1$. In two and three dimensions, where the critical exponents in percolation \cite{Stauffer and Aharony 1994} correspond to $-187/91+1=-1.05$ and $-1.18$, respectively; the relation~\ref{dim03} gives the values of $d_A=1.9~(d_S=2)$  and $d_A=2.46~(d_S=3)$, while the values presented by \cite{Stauffer and Aharony 1994} correspond to $d_A=91/48=1.89~(d_S=2)$  and $d_A=2.53~(d_S=3)$. Although for two and three dimensions it seems to work, for the critical dimension $d_S=6$, the relation gives a ratio $d_S/d_A=2$, while the reported one is $3/2$ \cite{Stauffer and Aharony 1994}.

\subsection{Avalanches in two and three dimensions}\label{ava 2 3 dim}

Above the critical dimension, different approaches have shown that the critical exponent of scale invariant avalanches corresponds to $-3/2$. Dissipation moves the cut-offs towards small avalanche size values, provoking a decrease in the mean value of avalanche sizes $\langle s \rangle$, thus breaking the criticality of the system. In two and three dimensions the situation is more complex: there is no universality in the values of the critical exponents, and dissipation can have different effects on the distribution, either changing the cut-off of the distribution \cite{Planet et al 2009} or decreasing the exponent of the power-law \cite{OFC 1992}.

\begin{figure}[b!]
\begin{center}
\vspace{-1.0cm}
\includegraphics{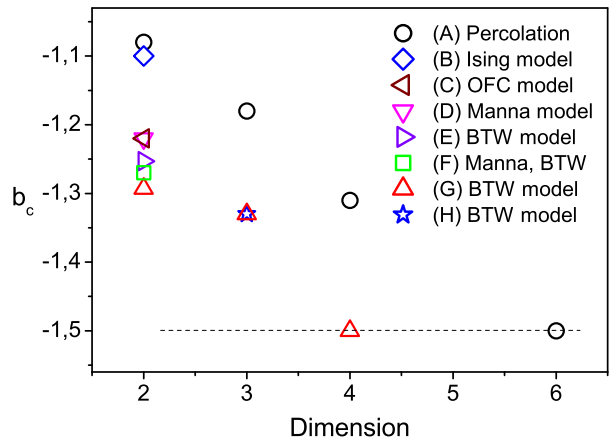}
\end{center}
\vspace{-1.9cm} \caption{\label{fig:exponents}
Critical exponents for several models with scale invariant avalanches in different dimensions: (A) \cite{Stauffer and Aharony 1994}, (B) Wolff algorithm (section~\ref{ising model}), (C) OFC model (conservative) \cite{OFC 1992}, (D) \cite{Pietronero et al 1994}, (E) \cite{Manna 1990}, (F) \cite{Chessa et al 1999}, (G) \cite{Lubeck and Usadel 1997}, (H) \cite{Tang and Bak 1998}.}
\end{figure}

Figure~\ref{fig:exponents} displays the critical exponents of the power-law distribution of avalanches for several models and in different dimensions. The models can be divided into three groups: percolation, Ising model, and slowly driven phenomena including the Olami-Feder-Christensen (OFC) \cite{OFC 1992} model, the BTW model \cite{BTW 1987} and the Manna model \cite{Manna 1990}. Percolation donates the highest exponent values, that decrease with the dimension until the critical dimension 6 where it reaches the value -3/2. The exponent value obtained in section~\ref{ising model} by the avalanche distribution in the Ising model (Wolff algorithm) is also displayed and it shows a value close to Percolation in two dimensions. The group of slowly driven phenomena in two dimensions shows values approximately in the interval ($-1.3, -1.2$), where there is also a report indicating universality with an exponent $-1.27$ between two of the presented models \cite{Chessa et al 1999}. For three dimensions there are much less studies, but some numerical results indicate a critical exponent equal to $-1.33$ for the BTW model. As mentioned earlier, the critical dimension for slowly driven phenomena is 4 \cite{Vespignani and Zapperi 1997}. Following the relation~\ref{dim03}, the different exponents must be related to the structure of the avalanches, in particular their fractal dimensions.

The lowest value of the critical exponent for the avalanches in slowly driven systems corresponds to $-3/2$ and takes place in dimension 4 (or superior). In three dimensions the results indicate a value equal to $-1.33$ and in two dimensions let us take $-1.27$ as the paradigm. However, the majority of real phenomena evolving through scale invariant avalanches present much lower exponents: earthquakes ($b=-2$), granular avalanches ($b=-1.6$) \cite{Ramos et al 2009}, superconducting vortices ($b=-1.6$) \cite{Bassler and Paczuski 1998, Altshuler and Johansen 2004, Altshuler et al 2004, Altshuler et al 2004b}, solar flares ($b=-1.8$) \cite{Hamon et al 2002}, subcritical fracture ($b=-1.5$) \cite{Santucci et al 2004}, and so on. Why do these phenomena ``move" apart from the critical values of their respective dimensions? Dissipation is one possible answer.

\begin{figure}[b!]
\begin{center}
\vspace{-0.8cm}
\includegraphics{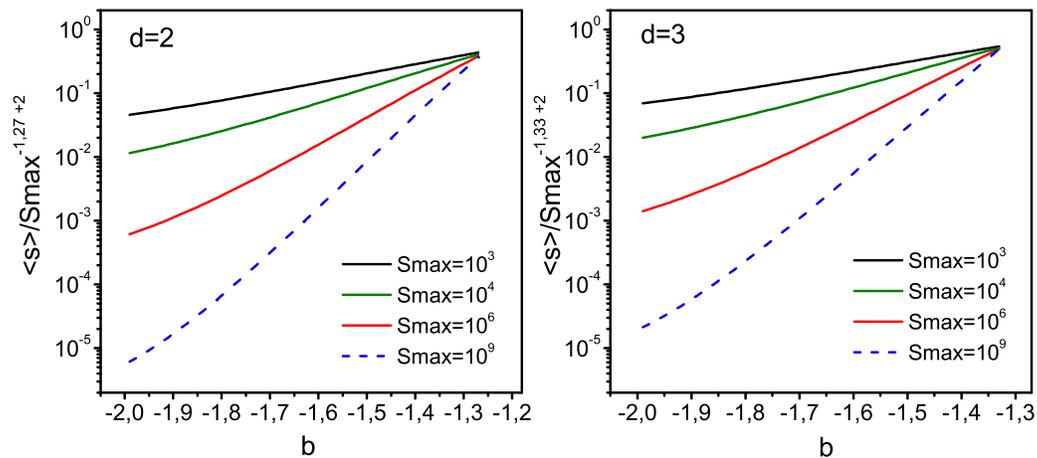}
\end{center}
\vspace{-4.8cm}\caption{\label{fig:dim2and3} Results for the condition of criticality $\langle s \rangle /S_{max}^{b_c+2}$ for slowly driven systems in two and three dimensions.}
\end{figure}

The Olami-Feder-Christensen (OFC) model of earthquakes is a nonconservative model that mimics the behavior of two tectonic plates, and is able to tune the exponent of the power-law distribution of avalanches by modifying the degree of dissipation in the system \cite{OFC 1992}. Recent and still unpublished results in granular piles (the continuation of \cite{Ramos et al 2009}) have shown the same tendency displayed by the OFC model: a decrease in the value of the exponent of the power-law distribution of avalanche sizes as dissipation increases. However, a quantitative relation between the dissipation and the exponent of the power-law is still lacking. Some reports have analyzed the effect of dissipation on the critical properties in the OFC model \cite{Carvalho and Prado 2000, Miller and Boulter 2002, Xia et al 2008} and also in a more general framework \cite{Bonachela and Munoz 2009}. The results show critical properties only in the conservative case; but again, a formalism linking the exponent of the power-law to the critical properties of the system is still missing.  From the eqs.~\ref{dim01} and~\ref{dim02} it is possible to get a condition of criticality for scale invariant avalanches:
\begin{equation}
\label{cond_crit}
\frac{\langle s \rangle}{S_{max}^{b_c+2}}\sim 1
\end{equation}
\noindent with the considerations of $b_c=-1.27$ and $b_c=-1.33$ for two and three dimensions respectively, in the case of slowly driven systems. The results of this condition of criticality are displayed in figure~\ref{fig:dim2and3}, where the system separates exponentially from the critical situation as the exponent $b$ decreases. For a given exponent the decrease grows linearly with the maximum avalanche size $S_{max}$.

In section~\ref{energy balance_avalanches} an energy balance limited the SOC in forbidding large values of $b$ (close to $-1$) for dissipative slowly driven system. Now the results indicate that as the exponent decreases, which in different systems is a consequence of the dissipations, the system loses its critical properties. The combination of both results restrict SOC to conservative and critical models, like the original BTW one. This argument is in agreement with recent results in avalanche prediction, which is the main subject of the next section.

\section{Towards Prediction and Control}

Several works have claimed the unpredictable character of phenomena evolving through scale invariant avalanches (earthquakes, granular piles, solar flares, stock markets, etc) as a consequence of their classification as critical systems \cite{Main 1999, Geller et al 1997}. However, in the last section it has been suggested that those systems are not critical, which is expressed in the small value of the exponent of their power-laws compared to the critical exponent at their respective dimensions. If they are not critical, they are, in principle, predictable; a fact that has been recently proved in both experiments and simulations.

\subsection{Predicting scale invariant avalanches}

\begin{figure}[t!]
\vspace{-0.6cm}
\includegraphics{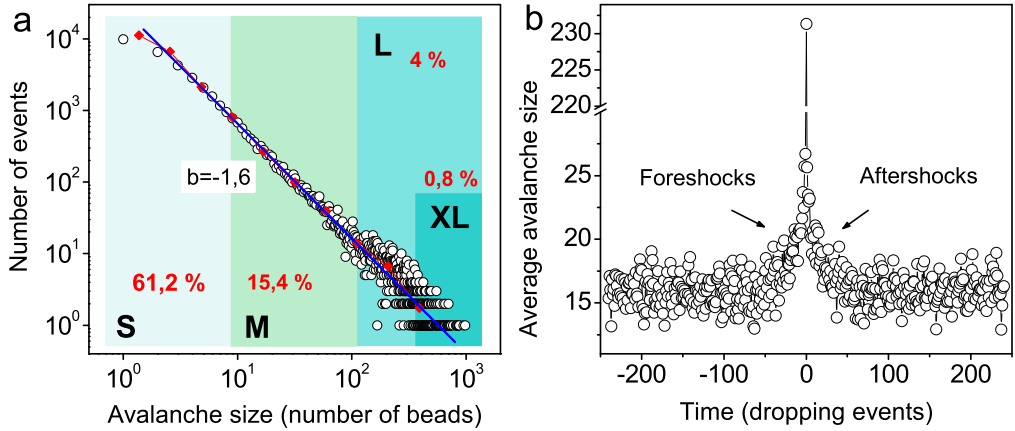}
\vspace{-4cm} \caption{\label{fig:experiment1} (a) Distribution of avalanche size for the granular experiment (circles). These points have been averaged with a logarithmic binning (diamonds) and they follow a power-law with an exponent $b=-1.6$. Avalanches have been classified as small (S), medium (M), large (L) and extra-large (XL). The percentage of occurrence of each type of avalanche is also displayed. (b) The average of the avalanche size around a large event (L) presents weak signs of foreshocks and aftershocks.}
\end{figure}

The experiment studies the dynamics of avalanches in a quasi-two-dimensional granular pile \cite{Ramos et al 2009}. It consists of a base of $60~cm$ row of randomly spaced $4~mm$ steel spheres, sandwiched between two parallel vertical glass plates $4.5~mm$ apart. The same steel beads are delivered one by one from a height of $28~cm$ above the base and at its center, resulting in the formation of a quasi-two-dimensional pile. The extremes of the base are open, leaving the beads free to abandon the pile. After a bead is delivered, the pile is recorded with a digital camera at a resolution of 21~pixels/bead-diameter, followed by the dropping of a new bead. One experiment contains more than $55000$ dropping events with a total duration of more than 310 hours. The first $4500$ events before the pile reaches a stationary state are not included in the statistics. The average number of beads in the pile is $3315$. The centers of all the particles for each image are found, and the size of an avalanche is defined as the number of beads that has moved between two consecutive dropping events. The distribution of avalanches follows a power-law with an exponent $b=-1.6$ (figure~\ref{fig:experiment1}a). Avalanches are classified as small, medium, large and extra-large; and the study focuses on predicting large and extra-large events. Foreshocks and aftershocks take place around large avalanches (figure~\ref{fig:experiment1}b). However, these signs are rather weak, and all the efforts to predict {\it when} a large avalanche is going to happen did not succeed: the analysis of the time series does not allow the prediction of large events.

\begin{figure}[t!]
\begin{center}
\vspace{0.5cm}
\includegraphics[width=5in]{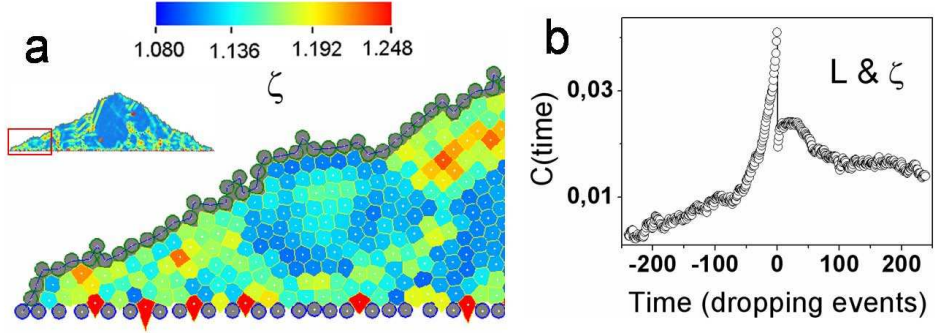}
\end{center}
\vspace{-0.5cm}
\caption{\label{fig:disorder} (a) Portion of the granular pile where the color of each voronoi cell represents the value of the shape factor. (b) Temporal cross-correlation between the large avalanches and the average shape factor.}
\end{figure}

As the position of the centers of each particle at every step of the experiment is known, different structural variables can be defined and their evolution followed during all the experiment, particularly in the neighborhood of a large avalanche. The shape factor $\zeta=C^2/4\pi S$, where $C$ is the perimeter and $S$ the area of each voronoi cell in which the structure has been divided (figure~\ref{fig:disorder}a), has been computed at each step of the experiment. $\zeta$ is equivalent to the local disorder and inversely proportional to the packing fraction of the pile. The temporal cross-correlation between the shape factor (spatially averaged) and the large avalanches read as:

\begin{equation}
\label{cross_corr}
C(t) = \frac{\sum s_b(t_i)\zeta(t_i+t)- \langle s_b(t_i) \rangle \langle \zeta(t_i) \rangle }
{\sqrt{\sum (s_b(t_i)- \langle s_b(t_i) \rangle)^2 \sum (\zeta(t_i)- \langle \zeta(t_i) \rangle)^2 }}
\end{equation}

\noindent where $s_b$ corresponds to the binary series of large (extra-large) avalanches, i.e., 1 if the avalanche is large (extra-large) and 0 otherwise. The results are shown in figure~\ref{fig:disorder}b. The continuous variation displayed by the average disorder of the pile before a large avalanche is very clear and approximately fifty steps before a large event, the average disorder continuously increases until the avalanche takes place. Then the pile reorganizes itself, but it gets trapped in an intermediate level of disorder. In the aftershocks zone, the disorder increases, and after that, the pile slowly evolves into more organized states. By using the information relative to the average disorder and the help of an extremely simple algorithm, it has been possible to reach up to $62\%$ and $64\%$ of success in the predictability of large and extra-large avalanches respectively. By changing the algorithm and adding more information these odds can be improved.

Very similar results have been obtained in a more realistic modification of the OFC model \cite{Ramos 2010, Ramos et al 2006}, consisting of a cellular automaton where the Burridge and Knopoff spring-block model has been mapped \cite{Burridge and Knopoff 1967}. The spring-block model consists of a two dimensional array of blocks on a flat surface. Each block is connected by means of springs with its four nearest neighbours, and in the vertical direction, to a driving plate which moves horizontally at velocity $v$. When the force acting on a block overcomes the static friction of the surface, the block slips. A redistribution of forces then takes place in the neighbors that eventually triggers new displacements. In our model \cite{Ramos et al 2006}, the force on each block is stored in a site of the lattice, and the static friction thresholds are distributed randomly following a Gaussian. If $20\%$ of the energy is lost in every redistribution of forces after one block slips, the distribution of avalanches resembles the Gutenberg-Richter law \cite{Christensen and Olami 1992}, showing a power-law with an exponent $b=-1.91$. The cross-correlation between the standard deviation of the force and the binary series of large avalanches donates a result very similar to the one presented in figure~\ref{fig:disorder}b for the granular experiment. The results are more pronounced for larger system sizes, indicating the possibility of prediction.

\subsubsection{Criticality. Good or bad for prediction?}

\begin{figure}[b!]
\begin{center}
\vspace{-0.8cm}
\includegraphics{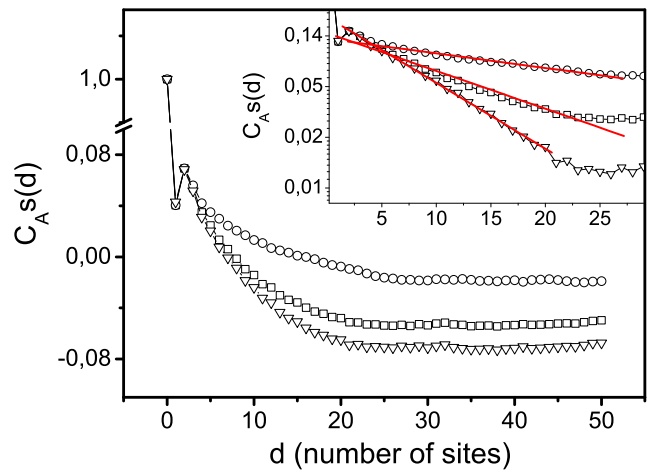}
\end{center}
\vspace{-0.8cm} \caption{\label{fig:tectono}
Averages of the spatial correlation function
calculated at 300 steps before a large event (superior curve), 300 steps after a large
event (inferior curve) and at equispaced intervals of 200 steps (middle curve), for
L=128. (Inset) From a linear fit in log-lin axes, correlation length values equal to 36,
14 and 7 sites can be extracted.}
\end{figure}

In a similar manner as in the Ising model, by following the eqs.~\ref{spatial corr_funct} and~\ref{corr length}, the correlation length $\xi$ has been calculated and is presented in figure~\ref{fig:tectono} \cite{Ramos 2010}. As it is expected considering the low value of the exponent of the power-law $b=-1.91$, the average over regular intervals has donated a low value of the correlation length $\xi = 14$ in a system size $L = 128$; thus the system is not critical. The correlation length has also been calculated averaging the correlation functions at certain steps before (and after) a large event. The value of $\xi$ calculated 300 steps before a large event corresponds to 36. This higher value seems to be an indication that in a temporal neighborhood preceding a large event, the system presents characteristics of a critical state, which has been denominated {\it ``temporal critical state"} \cite{Ramos 2010}. Therefore, if the average is performed over a ``moving" and relatively small time window, the correlation length will have small values most of the time, but will eventually reach values close to the system size in the neighborhoods of catastrophic events (temporal critical states). The existence of CSD in this temporary state could be extremely helpful in order to predict scale invariant avalanches \cite{Scheffer et al 2009}. Of course, inside the critical region it will not be possible to predict when a large avalanche will take place, but as the system size increases, the duration of this critical state must reduce. Concerning earthquakes, a report in 2009 has shown some measurements indicating a window duration of a few hours \cite{Manshour et al 2009}.

Our experiments and simulations present two major limitations: first, they are quasi-static, which makes impossible to analyze the existence of CSD. Second, the predictions are based on the internal structure, and although there are some studies claiming the possibility of predicting earthquakes following a similar idea \cite{Crampin et al 2008, Crampin and Gao 2010}, the results are rather limited. However, there is an indirect sign indicating the existence of CSD in our experiment: the increase of disorder related to a reduction of the packing fraction in the system. In the percolation problem of the forest fires, the average termination time of the fire diverges at the percolation threshold \cite{Stauffer and Aharony 1994}. At this critical point, the structure is connected in a complex way and we can imagine the spatial scenario as a labyrinth that the fire has to cross. In our system the increase of disorder related to a reduction of the packing fraction creates also a complex path, that can eventually slow down the ``transfer of information" between different part of the system. In a more quantitative manner, in the jamming transition of a granular system, a decrease in the packing fraction provokes an increase of the soft modes \cite{Wyart et al 2005} which are modes of very low frequency. This shift to low values in the frequency is equivalent to an increase of the correlation time (which is the signature of a CSD).

In the latest years, the possibility of continued measurements, storage and data analysis of several networks of seismometers and accelerometers has brought promising studies based mainly on the analysis of the cross-correlation function of the seismic noise \cite{Brenguier et al 2008, Brenguier et al 2008b}. However, no precursors of large events have been found with this method; and probably the answer can also be found in the jamming transition, because the soft modes do not have any influence on the elastic properties of the medium.

\subsection{The origin of Scale invariant avalanches}

The origin of the temporal scale invariance in nature has been a question of debate for more than two decades \cite{Bak 1997, Vespignani and Zapperi 1998, Reed 2002}. Being able to group many different phenomena (earthquakes, granular avalanches, solar flares, stock markets, etc) into the same kind of dynamics, where catastrophic events and small events are explained by the same principles, was an extraordinary achievement of the SOC. However, the axiomatic manner in which the base of the theory was introduced, related to the existence of a critical state as an attractor of the dynamics, set SOC as a theory to be
proved more than as a theory to develop. This has provoked that several relevant questions concerning scale invariant avalanches have never been formulated in a direct manner; the influence of the dissipation and the influence of the disorder in the ``self-organization" are examples of these questions.

Although there is not quantitative results about the two proposed questions, two decades of work on the subject have left some hints about them: the effect of the dissipation on modifying the exponent of the power-law has already been discussed in section~\ref{ava 2 3 dim};  and now let us analyze  the influence of the disorder in the ``self-organization" towards scale invariant avalanches. Two cases are going to be discussed: a granular pile and an earthquake model. In these situations (as well as all the other SOC systems) an energy gap controls the limits of the dynamics, i.e., the largest amount of dissipated energy. In the granular pile two angles define the energy gap: the subcritical angle and the supercritical angle \cite{Daerr and Douady 1999}. In ideal conditions (little disorder and large friction) a trivial periodic behavior rules the dynamics \cite{Held et al 1990, Jaeger et al 1990, Rosendahl et al 1993, Rosendahl et al 1994}, charging and discharging the energy gap. In the earthquake model \cite{OFC 1992, Ramos et al 2006} there is some elastic energy stored in every site of a lattice, which is limited by local thresholds related to a static friction coefficient. The largest possible avalanche happens when all the sites have reached their thresholds and, with the trivial condition of a flat distribution of thresholds, a trivial periodic behavior also rules the dynamics.

\begin{figure}[h!]
\begin{center}
\vspace{-2.0cm}
\includegraphics{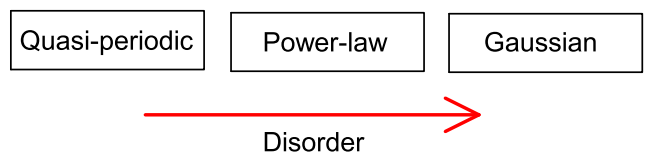}
\end{center}
\vspace{-4.0cm} \caption{\label{fig:origin}
The disorder is able to modify the type of distribution in a system with an energy gap.}
\end{figure}

If the structural disorder in the granular pile is relevant, the periodicity is broken and avalanches become temporally uncorrelated with sizes distributed following a power-law \cite{Frette et al 1996, Altshuler et al 2001, Aegerter et al 2003}. In the case of the earthquake model, a more realistic Gaussian distribution of static friction thresholds will be sufficient in order to obtain a power-law distribution of events \cite{Ramos et al 2006}; however, some signs of periodicity (proportional to the dissipation) are still present in the dynamics. By introducing more disorder (a Gaussian distribution of the values of the dissipation), the periodicity is broken while the avalanche size distribution remains as a power-law. Nevertheless, increasing the values of the standard deviation of the Gaussian distribution will lead to the removal of the power-law behavior \cite{Ramos et al 2006}. These simulations were focused on the SOC behavior; thus no more disorder was introduced after the rupture of the scale invariance; though it is evident that if a disorder is artificially increased, a Gaussian distribution behavior must be obtained. A simple example will be not respecting the separation of scales (drive and dissipation) required for a scale invariant behavior, and to drive continually a granular pile. The result will be a continuous flow and a Gaussian behavior will be obtained.

Predicting scale invariant avalanches in natural phenomena (particularly earthquakes), is one of the biggest challenges of today's science. However, predicting catastrophic avalanches it is not the final solution to this problem. The question to address is more practical: Is it possible to control scale invariant avalanches? The simplest solution in order to break the scale invariance is the {\it early triggering}, which is currently used in snow avalanches and in numerical models \cite{Cajueiro and Andrade 2010} to avoid large accumulations leading to catastrophic events. Understanding the role of dissipation, disorder, and other factors linked to the organization of a system into a dynamics of scale invariant avalanches, can be essential in the future development of tools leading to the control of catastrophic avalanches.

\section{Conclusion}

The essential role of the exponent value of the power-law distribution of avalanche sizes has been discussed. The exponent value controls the proportion between small and large events in the dynamics. For an exponent $b=-1$, for example, the probability of large and small events is the same (considering a logarithmic resolution). As the exponent value decreases (increases in absolute value), smaller events have more weight in the dynamics. This is also presented in the value of $\langle s  \rangle$, related to the energy balance in the system, and forbidding exponents close to $b=-1$ for slowly driven systems. The exponent value also controls the critical properties of the system through the {\it condition of criticality} $\langle s \rangle /S_{max}^{b_c+2}$ where $b_c$ corresponds to the critical exponent for slowly driven systems, with $b_c \sim -1.27$ and $b_c \sim -1.33$ in two and three dimensions respectively.

The causes, consequences, as well as some history of the logarithmic scale have also been presented. The logarithmic scale is the only scale that respects the scale invariance during the measurement. The fact that an integration is performed during the measurement changes the value of the exponent in $+1$ unit, which has provoked some confusion in the interpretation of the distribution of earthquakes. However, it has been shown that a power-law with an exponent $b=-2$, and the consideration of one earthquake occurring every $0.21$ seconds, fits quite well the real data.

Simulation on a well established critical system (the Ising model) has revealed that power-law distributed avalanches are not a necessary condition in order to classify a system as {\it critical}, but different kinds of distributions can rule the avalanche behavior of an equilibrium system at the critical point: A ``Metropolis" algorithm displayed avalanches distributed following a Gaussian, while the ``Wolff" algorithm presented a scale invariant dynamics with a critical exponent  $b=-1.1$. Different experiments have shown different kind of dynamics.

Real scale invariant phenomena (earthquakes, granular piles, solar flares, etc) normally present exponent values far from the critical ones, and simulations and table experiments suggest that dissipation is the cause of these low values. As the exponent moves apart from the the critical value, the {\it condition of criticality} indicates that the system is loosing the critical properties, thus it becomes, in principle, predictable. Prediction of scale invariant avalanches has been achieved both in a quasi-two-dimensional granular pile ($b=-1.6$) and in an earthquake model ($b=-1.91$). The correlation length $\xi$ has been calculated in the earthquake model showing a low value when the correlation function is averaged over the whole dynamics (or non correlated events). However, if $\xi$ is calculated averaging the correlation function over a small time window, it presents small values most of the time, but eventually it reaches values proportional to the system size preceding catastrophic events, a situation denominated {\it ``temporal critical state"}. The analysis of this temporal state can eventually be used to forecast real catastrophic avalanches.

The influence of dissipation and disorder in the ``self-organization" of scale invariant dynamics have been discussed in a qualitative manner. However, quantitative studies are still missing. These  studies can eventually have an impact in the prediction and control of real catastrophic avalanches.

\section{Acknowledgments}

I am pleased to thank N. Arnesen and O. Duran for proofreading the manuscript; E. Altshuler, S. Ciliberto, O. Duran, P. Holdsworth, K. J. M{\aa}l{\o}y, T. Roscilde, S. Santucci and L. Vanel for useful discussions; J.-F. Pinton for sharing some experimental data; R. Planet and S. Santucci for providing the figure~\ref{fig:stef}; and C. Creton and A. Lindner for a lot of support during the writing of this chapter. I am also pleased to thank Henrik Jeldtoft Jensen, who reviewed the manuscript and made very useful comments. This work has been financed by the European project MODIFY.




\label{lastpage-01}

\end{document}